\documentclass[showpacs,aps,prb,twocolumn]{revtex4}

\usepackage{graphicx}
\usepackage{amsmath}
\usepackage{subfigure}
\usepackage{dcolumn}

\begin{document}


\title{
A first-principles characterization of the structure and electronic
structure of $\alpha$-S and Rh-S chalcogenides
}

\author{Oswaldo Di\'eguez}

\affiliation{Institut de Ci\`encia de Materials de Barcelona (ICMAB-CSIC),
             Campus UAB,
             08193 Bellaterra, Spain}

\author{Nicola Marzari}

\affiliation{Department of Materials Science and Engineering,
             Massachusetts Institute of Technology,
             77 Massachusetts Avenue,
             Cambridge MA 02139, USA}


\begin{abstract}
We have used first-principles calculations to study the structural,
electronic, and thermodynamic properties of the three known forms of Rh-S
chalcogenides:
Rh$_2$S$_3$, Rh$_3$S$_4$, and Rh$_{17}$S$_{15}$.
Only the first of these materials of interest for catalysis had been 
studied previously within this approach.
We find that Rh$_{17}$S$_{15}$ crystallizes in a {\em Pm}\={3}{\em m}
centrosymmetric structure, as believed experimentally but never confirmed.
We show band structures and densities of states for these compounds.
Finally, we also investigated the ground state structure of solid sulfur
($\alpha$-S), one of the few elements that represents a challenge for full
first-principles calculations due to its demanding 128-atom unit cell and
dispersion interactionis between S$_8$ units.
\end{abstract}

\date{\today}

\pacs{
61.50.-f 
71.20.-b 
}

\maketitle


\section{Introduction}

There is currently a growing interest in rhodium sulphide materials because of
the potential of transition-metal chalcogenides as catalysts in fuel cells,
substituting expensive and scarce platinum.\cite{Vante1986,Vante1987}
Rhodium sulfide possesses promising characteristics, from significant activity
in acid electrolytes to selectivity towards oxygen reduction reaction in large 
concentrations of methanol.\cite{Ziegelbauer2006}
Recently, a commercially available carbon-supported Rh$_x$S$_y$ electrocatalyst
has shown a superior stability and performance over a
sample of other chalcogenides, while being free of the toxicity problems 
intrinsic to those containing selenium.\cite{Ziegelbauer2008}
Other than in fuel cells, rhodium sulfide is useful as a catalyst in 
situations like efficient electrolysis of aqueous HCl solutions for the
industrial recovery of high-value chlorine gas.\cite{Gulla2007}

The Rh$_x$S$_y$ material used for electrocatalysis contains a mixture of the
three phases of rhodium sulfide that are positively known to exist:
Rh$_2$S$_3$, Rh$_3$S$_4$, and Rh$_{17}$S$_{15}$.
Despite of the interest on their catalytic properties, not much is known about
these materials, in part due to their complex crystal structures
(with 20, 42, and 64 atoms per unit cell, respectively).
In this situation, we have used first-principles calculations as an effective
tool to extract valuable information about these systems.
In the same spirit, and as a way to validate our methodology, we also 
investigated the structural properties of the ground state of pure sulfur 
($\alpha$-S).
Sulfur is one of the very few elements for which full first-principles 
relaxation calculations have not been performed (boron being another element
known for the difficulties that its crystalline ground state 
presents\cite{Ogitsu2009JACS}).

The rest of this paper is organized as follows.
We describe the method we used in Section II, present our main results in 
Section III, and summarize our work in Section IV. 


\section{Methodology}

For our calculations, we have used density-functional theory 
(DFT)\cite{Hohenberg1964} in the Kohn-Sham framework\cite{Kohn1965}, as 
implemented in the {\sc PWscf} program of the {\sc Quantum-ESPRESSO} 
package.\cite{espresso}
For the exchange correlation functional we employed the generalized-gradient
approximation (GGA) functional developed by Perdew, Burke, and
Ernzerhof (PBE),\cite{Perdew1996} since it is known that GGA gives somewhat
better results than the simpler local-density approximation
when describing the structural properties of this kind of transition-metal
sulfides.\cite{Raybaud1997}
The electron-ion interactions are described through the use of
ultrasoft pseudopotentials.\cite{Vanderbilt1990} 
These pseudopotentials were generated with scalar-relativistic calculations
in the $4d^7 5s^2$ configuration for Rh, and in the $3s^2 3p^4$ configuration
for S.
More specific details about the pseudopotentials can be found in
Ref.\ \onlinecite{pseudos}.

The Kohn-Sham equations\cite{Kohn1965} were solved self-consistently by using 
a plane-wave basis set, and the integrations in reciprocal space were
performed using Monkhorst-Pack grids.\cite{Monkhorst1979}
The amount of plane-waves and grid points used depended on the system studied,
and more details are given in the next section.


\section{Results}

As a way to control the numerical approximation involved in our methodology,
we studied the properties of the isolated crystalline elements Rh and S, and
we compared the results with the well known experimental ones. 
In the process we have done what we believe are the first DFT calculations 
with full relaxation of the 128-atom unit cell of the most stable phase of
S at ambient conditions.
Then we then applied the same methodology to compute the structural and
electronic properties of three Rh-S compounds:
Rh$_2$S$_3$, Rh$_3$S$_4$, and Rh$_{17}$S$_{15}$.

\subsection{Pure elements}

\subsubsection{Rhodium}

Rhodium is a precious metal that crystallizes in a face-centered cubic lattice
with one atom per unit cell, corresponding to space group \#225 ($Fm\bar{3}m$).

Our DFT calculations for crystalline Rh use a plane-wave cutoff of
30 Ry and a $4 \times 4 \times 4$ k-point mesh.
We obtained a optimized lattice parameter of 3.862~\AA~(Rh-Rh distance of
2.731~\AA) and a bulk modulus of 258 GPa.
This shows a good agreement with experimental values (3.80~\AA~and 270 GPa,
respectively), and with other theoretical calculations (see
Ref.~\onlinecite{Pozzo2007} and references therein).

\subsubsection{Sulfur}

Sulfur can crystallize in around thirty different phases depending on pressure
and temperature, a number that makes it the chemical element with the highest
number of allotropes in the periodic table.\cite{Steudel2003} 
At ambient pressure it forms structures with a very small packing factor.
Twenty of those are composed of molecular rings with six to twenty sulfur 
atoms each.
There also exist polymeric structures formed by molecular chains.
As pressure increases the structures get more closely packed, and in the
megabar regime sulfur becomes a metal with a superconducting temperature of
17 K.\cite{Struzhkin1997}

\begin{figure}
\subfigure[]{\includegraphics[width=2.5in]{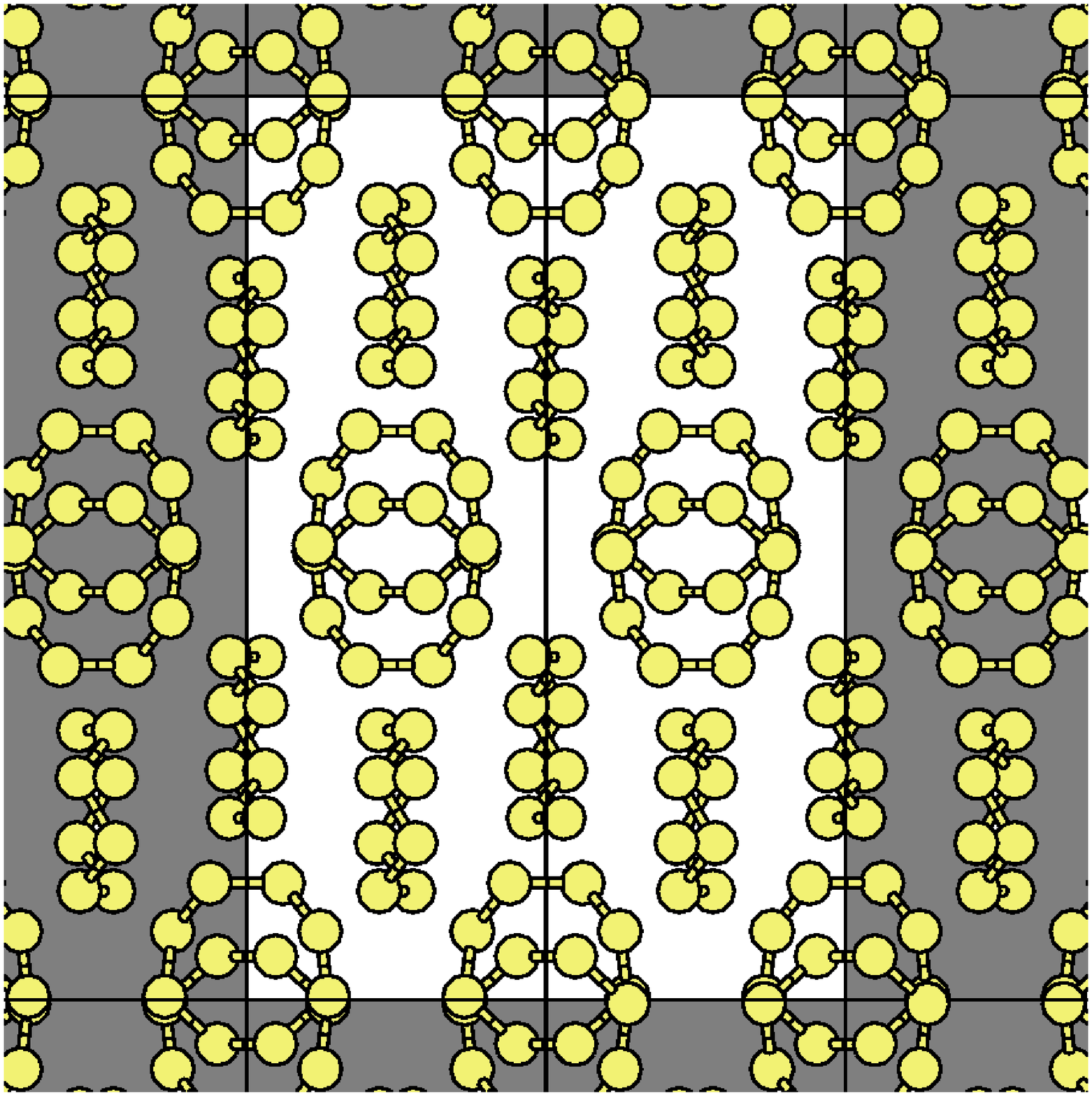}}
\subfigure[]{\includegraphics[width=2.0in]{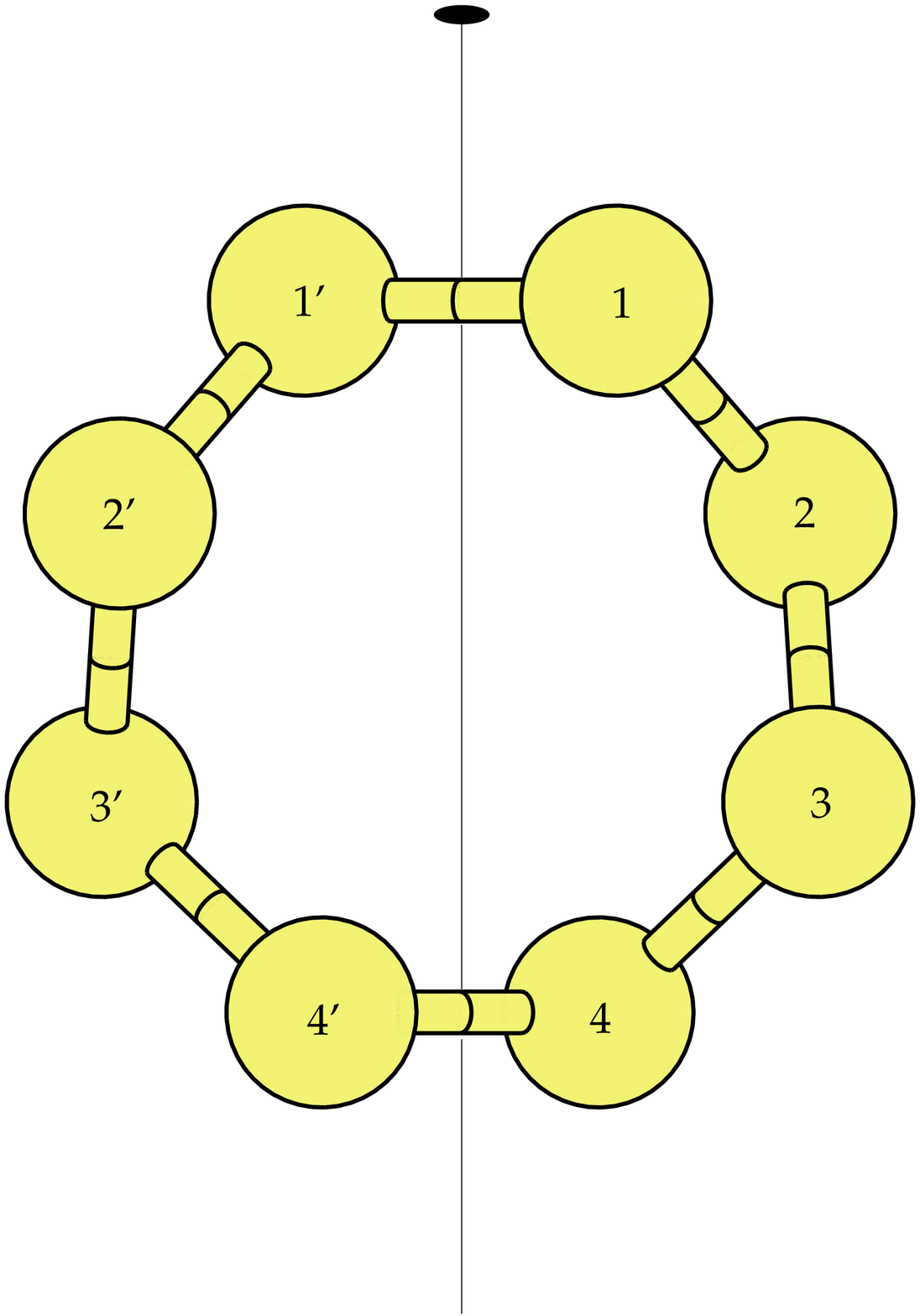}}
\caption{
(a) Structure of $\alpha$-S viewed from the [110]
axis, with the [001] axis along the lenght of the page; the unit cell (white 
background) contains sixteen S$_8$ molecules.
(b) One of the S$_8$ molecules, showing the $C_2$ axis along the [001]
direction of the crystal; the labels on the atoms will be used in the 
text to identify them.}
\label{fig_Sstructure}
\end{figure}

At ambient conditions the most stable phase of sulfur is $\alpha$-S, in which
$S_8$ molecules crystallize in the orthorrombic space group \#70 ($Fddd$). 
Figure \ref{fig_Sstructure} (a) shows the unit cell of $\alpha$-S,
containing sixteen S$_8$ molecules.
Due to the interactions with its neighbors, the original $D_{4d}$ symmetry 
of each S$_8$ molecule is lost, and instead it has $C_2$ symmetry with the
rotation axis along the $c$-axis of the solid, as shown in 
Fig.\ \ref{fig_Sstructure}(b).

We have performed DFT calculations for $\alpha$-S
using a plane-wave cutoff of 30 Ry and the $\Gamma$ point for the reciprocal
space integrals.
The plane-wave cutoff was high enough as to provide results that were converged
better than 1 part in 1000 for the bond length, bond angles, and dihedral 
angles of the S$_8$ molecule.
We tested convergence with respect to k-point sampling by doing a calculation
at fixed experimental lattice parameters using four grid points, including 
the $\Gamma$ point and three more points at the boundary of the Brillouin zone.
After optimization the geometrical features of the atomic arrangement did not
show any important change, with the bond lengths varying less than $0.4\%$,
and the angles varying less than $0.8\%$.
Table \ref{tab_S} shows the structural results we have obtained once the
atoms were relaxed (using both in the fixed experimental cell vectors, and free
cell vectors) and a comparison with experimental data.
The lengths of the cell vectors are larger than the experimental results by
$22.0\%$, $13.8\%$, and $6.4\%$.
However, the S-S bonds in every S$_8$ ring are better than 1\% in agreement
with experiments.
Moreover, these rings show very similar geometrical features when the cell
used is the smaller experimental one, or when the computations are 
performed on an isolated ring.
These facts are one more illustration of the inability of functionals
like PBE to deal with non-local van der Waals interactions like the ones
between S$_8$ rings in orthorhombic sulfur. 
Very promising developments within DFT to deal with systems where these
interactions are important exist,\cite{Dion2004PRL} and they have been made
computationally effective very recently.\cite{RomanPerez2009}

\begin{table*}
\caption{Lattice parameters, bond lenghts, bond angles, and dihedral angles
         for $\alpha$-S as obtained experimentally,\protect\cite{Rettig1987}
         and computed in this work.
         The three theory columns refer to a fully relaxed cell,
         to a cell with fixed experimental lattice vectors,
         and to the relaxation of the isolated S$_8$ molecule.}
\begin{center}
\begin{tabular}{lcccc}
\hline \hline
  & Experiment  & \multicolumn{3}{c}{Theory (this work)} \\
  & $\alpha$-S  &  $\alpha$-S  & $\alpha$-S (fixed cell) &  Isolated S$_8$   \\
\hline
$a$                 
  &  10.4646 \AA     &  12.772 \AA     &  (10.4646 \AA)  &       -         \\ 
$b$                 
  &  12.8660 \AA     &  14.639 \AA     &  (12.8660 \AA)  &       -         \\
$c$                 
  &  24.4860 \AA     &  26.045 \AA     &  (24.4860 \AA)  &       -         \\
\hline
S(1)-S(1')          
  &  2.055 \AA       &  2.070 \AA      &  2.059 \AA      &  2.070 \AA     \\
S(1)-S(2)          
  &  2.054 \AA       &  2.071 \AA      &  2.067 \AA      &  2.070 \AA     \\
S(2)-S(3)           
  &  2.057 \AA       &  2.070 \AA      &  2.066 \AA      &  2.070 \AA     \\
S(3)-S(4)           
  &  2.056 \AA       &  2.072 \AA      &  2.069 \AA      &  2.070 \AA     \\
S(4)-S(4')          
  &  2.050 \AA       &  2.068 \AA      &  2.061 \AA      &  2.070 \AA     \\
Mean S-S            
  &  2.055 \AA       &  2.070 \AA      &  2.066 \AA      &  2.070 \AA     \\
\hline
S(1')-S(1)-S(2)     
  &  109.02$^\circ$  &  109.4$^\circ$  &  110.3$^\circ$  &  109.3$^\circ$ \\
S(1)-S(2)-S(3)      
  &  108.01$^\circ$  &  109.5$^\circ$  &  109.0$^\circ$  &  109.3$^\circ$ \\
S(2)-S(3)-S(4)      
  &  107.39$^\circ$  &  109.1$^\circ$  &  108.5$^\circ$  &  109.3$^\circ$ \\
S(3)-S(4)-S(4')     
  &  108.42$^\circ$  &  109.2$^\circ$  &  109.0$^\circ$  &  109.3$^\circ$ \\
Mean S-S-S          
  &  108.2$^\circ$   &  109.3$^\circ$  &  109.2$^\circ$  &  109.3$^\circ$ \\
\hline
S(2')-S(1')-S(1)-S(2)
  &  95.26$^\circ$   &  96.8$^\circ$   &  94.5$^\circ$   &  97.2$^\circ$  \\
S(1')-S(1)-S(2)-S(3) 
  &  98.12$^\circ$   &  97.1$^\circ$   &  96.4$^\circ$   &  97.2$^\circ$  \\
S(1)-S(2)-S(3)-S(4)  
  & 100.81$^\circ$   &  97.1$^\circ$   &  98.8$^\circ$   &  97.2$^\circ$  \\
S(2)-S(3)-S(4)-S(4') 
  &  98.84$^\circ$   &  97.2$^\circ$   &  98.2$^\circ$   &  97.2$^\circ$  \\
S(3)-S(4)-S(4')-S(3')
  &  96.94$^\circ$   &  97.6$^\circ$   &  96.9$^\circ$   &  97.2$^\circ$  \\
Mean S-S-S-S        
  &  98.5$^\circ$    &  97.2$^\circ$   &  97.3$^\circ$   &  97.2$^\circ$  \\
\hline \hline
\end{tabular}
\label{tab_S}
\end{center}
\end{table*}

\subsection{Rh-S compounds}

The rhodium sulfide compounds seem to have been investigated first by Juza,
H\"ulsmann, Meisel, and Blitz, who found several phases for these
chalcogenides in 1935.\cite{Juza1935}
One of them, believed to be ``Rh$_9$S$_8$'', turned out to be a
superconductor with a transition temperature of 5.8 K.\cite{Matthias1954}
Almost thirty years later, Geller showed that its composition was actually
Rh$_{17}$S$_{15}$, and that 
this is a cubic material belonging to one of three possible space
groups: {\em Pm}\={3}{\em m} (\#221, $O_h$), {\em P}\=43{\em m} (\#215, $T_d$),
or {\em P}432 (\#207, $O$).\cite{Geller1962}
It was not possible at the time to make a choice of the most probable
of the three space groups to which the crystal might belong. 
No further attempts seem to have been made in order to pursue this
question.
Most of the literature published afterwards refers to Rh$_{17}$S$_{15}$ as 
the centrosymmetric structure {\em Pm}\={3}{\em m}, since that is the one
for which lowest standard errors were obtained in the crystallographic 
analysis, even if Geller himself writes that ``this in itself does not 
necessarily mean that it is the most probable one''.\cite{Geller1962} 
A mineral with composition Rh$_{17}$S$_{15}$ receives the name of 
prassoite or miassite.

Shortly after Rh$_{17}$S$_{15}$ was characterized, the structure of 
Rh$_2$S$_3$ was also determined.\cite{Parthe1967}
This is a diamagnetic semiconductor with orthorhombic space group {\em Pbcn} 
(\#60, $D_{2h}$).
The authors also pointed out the impossibility of preparing a hypotetical 
``RhS$_2$'' phase that had been reported previously, stating that attempts to
do so led to a two-phase mixture containing elementary S and Rh$_2$S$_3$. 
Rh$_2$S$_3$ is found as a mineral called bowieite, where part of the Rh atoms
are substituted by Pt and Ir.

The original work by Juza {\em et al.}\ described also a rhodium sulfide
of the form Rh$_3$S$_4$.\cite{Juza1935}
However, it was not until 2000 that the characterization of this structure
was published,\cite{Beck2000} after Rh$_3$S$_4$ was synthetized at high
temperature.
This is a monoclinic structure with space group $C2/m$ (\#12, $C_{2h}$).
It corresponds to a metal that shows a weak paramagnetism down to
4.5~K.\cite{Beck2000}
Recently, the mineral kingstonite of ideal composition Rh$_3$S$_4$
has been found by the Bir Bir river in Ethiopia, although again with Ir and Pt
substituting to some degree Rh atoms.\cite{Stanley2005}

In the following we apply a first-principles method to investigate the
three materials described (Rh$_2$S$_3$, Rh$_3$S$_4$, and Rh$_{17}$S$_{15}$). 
These are, as far as we know, all the rhodium sulfides that have been 
characterized experimentally. 
Only Rh$_2$S$_3$ had been previously studied from
first-principles.\cite{Raybaud1997}

\subsubsection{Structures}

We relaxed the unit cell vectors and atomic positions of the three materials
studied.
After careful tests, we used a plane-wave cutoff of 40 Ry and 
Monkhorst-Pack grids of $2 \times 2 \times 2$.
This implied using eight non-equivalent k-points for Rh$_2$S$_3$ and
Rh$_3$S$_4$, and four non-equivalent k-points for Rh$_{17}$S$_{15}$.

\begin{figure}
\subfigure[]{\includegraphics[width=2.8in]{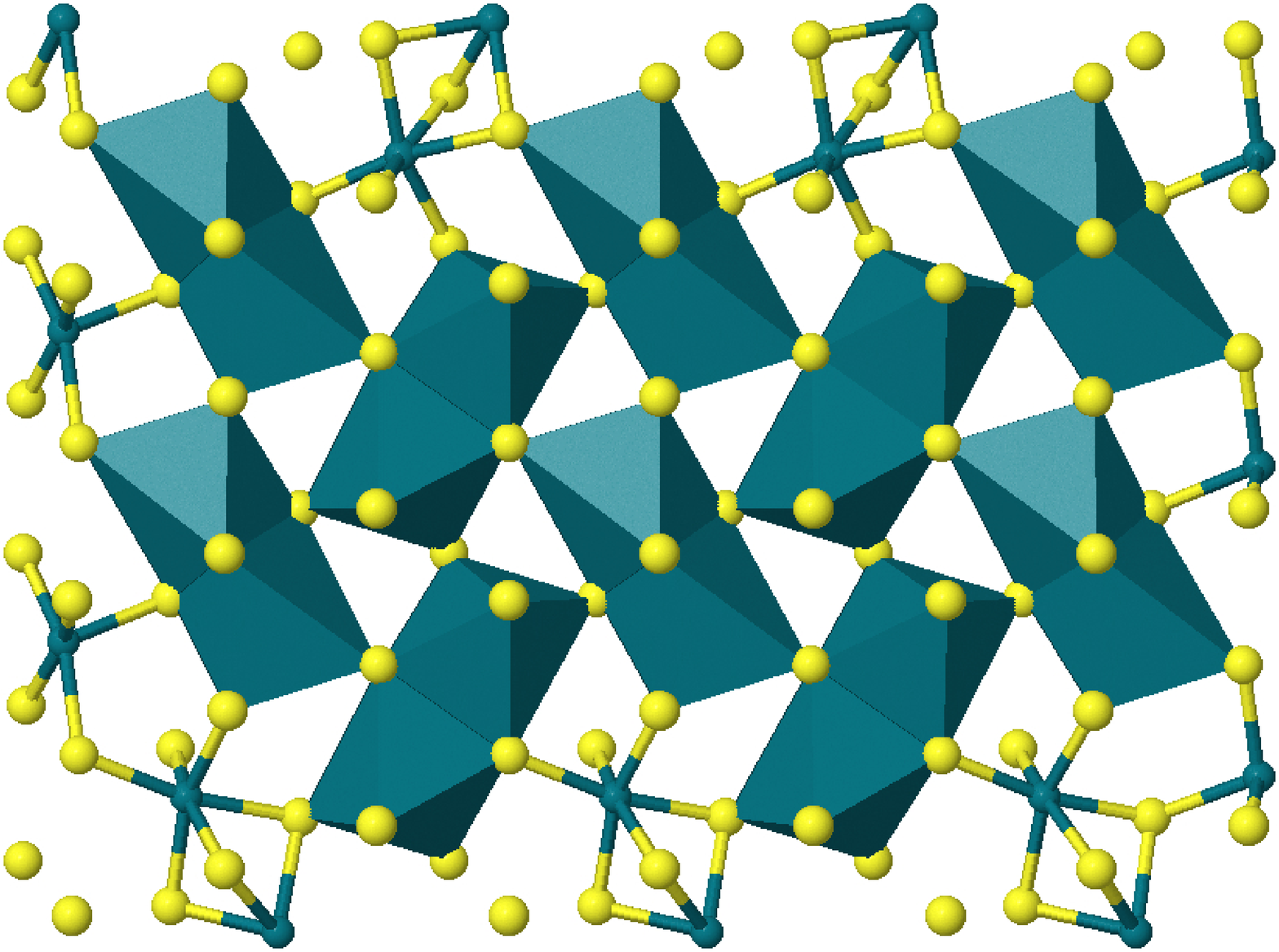}}
\subfigure[]{\includegraphics[width=2.8in]{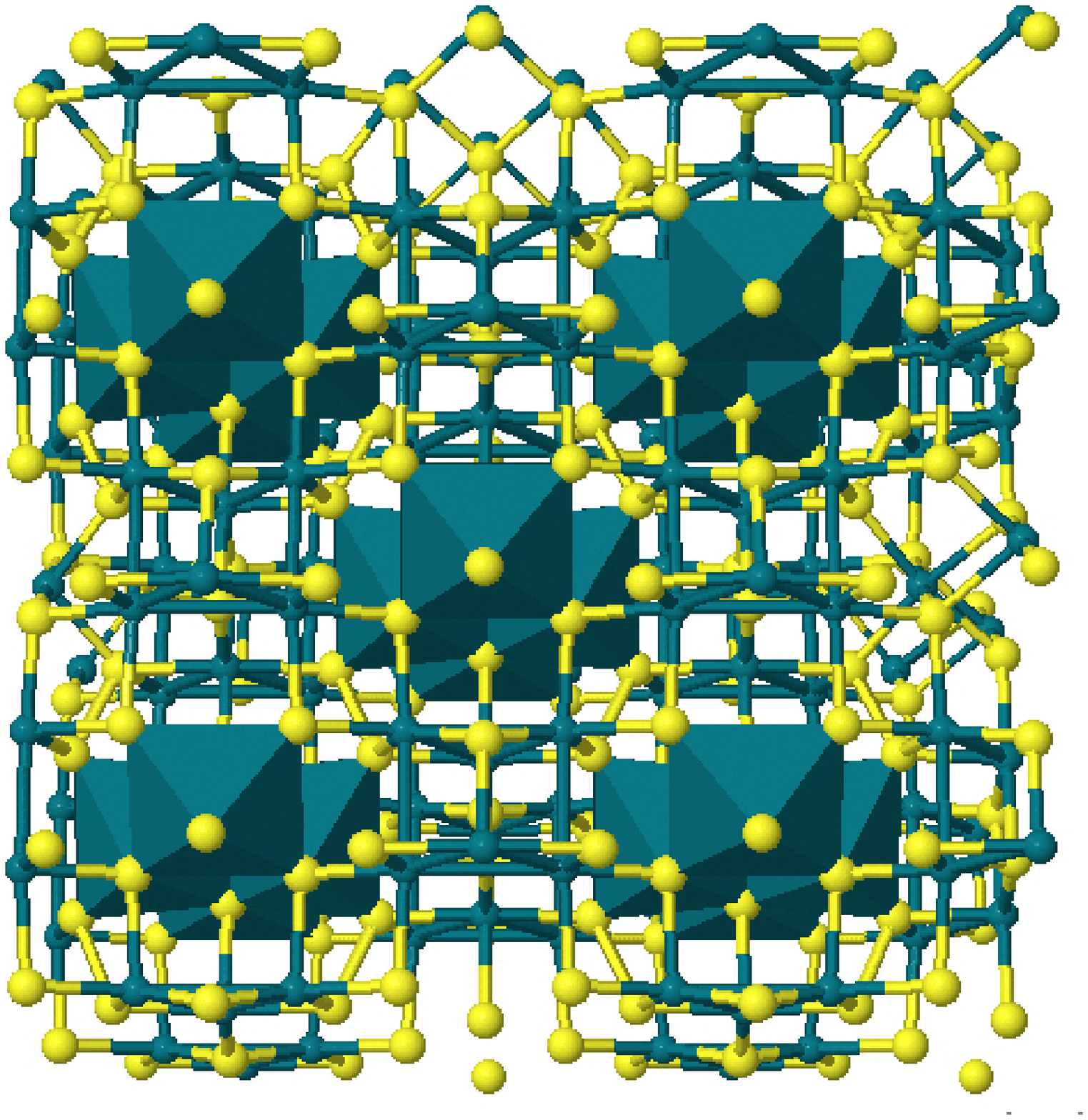}}
\subfigure[]{\includegraphics[width=2.8in]{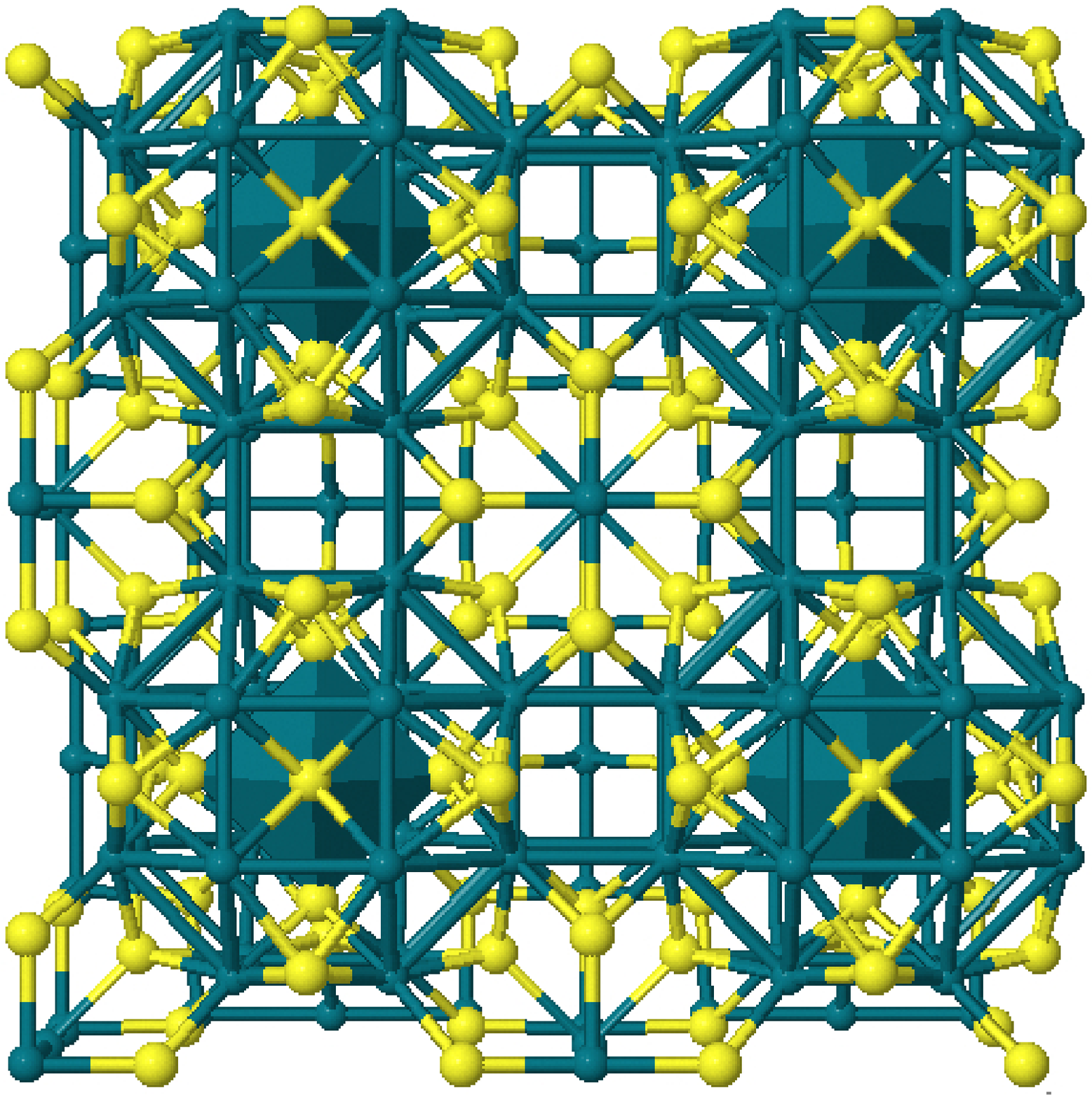}}
\caption{
Atomic structures obtained in this work for Rh-S materials:
(a) $3 \times 2 \times 3$ supercell for orthorhombic Rh$_{2}$S$_{3}$,
showing one of the two layers of face-sharing RhS$_6$ octahedra present in
the primitive 20-atom unit cell;
(b) $2 \times 2 \times 2$ supercell for monoclinic Rh$_{3}$S$_{4}$, showing
the octahedra surrounding two of the four inequivalent Rh atoms, and the
bonds between the other two types of Rh atoms;
(c) $2 \times 2 \times 1$ supercell for cubic Rh$_{17}$S$_{15}$, showing
the octahedra surrounding one of the four inequivalent Rh atoms.}
\label{fig_cells}
\end{figure}

Figure~\ref{fig_cells}(a) shows the atomic arrangement we obtained for 
orthorhombic Rh$_2$S$_3$.
The structural information for this material is summarized in
Table~\ref{tab_Rh2S3_struc}.
In this crystal there is only one inequivalent Rh atom, that is surrounded by
six S atoms located at the vertices of a distorted octahedron. 
There are two inequivalent S atoms, both surrounded by four Rh atoms forming a
distorted tetrahedron.
The whole structure can be described as an alternating stacking sequence of 
octahedron pair layers along [010], where each layer is related to a 
neighboring one by a $b$-glide reflection.
The cell parameters computed agree very well with experiment, 
the errors being 1.4\% ($a$), 1.1\% ($b$), and 1.2\% ($c$).
Apart from slightly larger bonds, the PBE approximation gives excellent
geometric predictions
for this compound, finding exactly the same patterns of distortion that are
found experimentally. 
The closest Rh-Rh distance is 3.208~\AA, substantially larger than the Rh-Rh
bond in pure Rh (2.731~\AA).
Our findings are very close to those of Raybaud and 
coworkers\cite{Raybaud1997}, who used a similar methodology.

\begin{table}
\caption{Lattice parameters, Wickoff atomic positions, and nearest-neighbor
distances in Rh$_2$S$_3$ (space group 60, $Pbcn$).
The experimental data is from Ref.~\protect\onlinecite{Parthe1967}.}
\begin{center}
\begin{tabular}{llcc}
\hline \hline
    & &  Experiment & Theory (this work) \\
\hline
    & $a$      & 8.462 \AA & 8.577 \AA \\
    & $b$      & 5.985 \AA & 6.049 \AA \\
    & $c$      & 6.138 \AA & 6.211 \AA \\
\hline
    Rh(1) & $8d$  & 0.10645 & 0.1071 \\
          &       & 0.2517  & 0.2513 \\
          &       & 0.0338  & 0.0331 \\
    S(1)  & $8d$  & 0.1518  & 0.1512 \\
          &       & 0.3906  & 0.3895 \\
          &       & 0.3930  & 0.3941 \\
    S(2)  & $4c$  & 0       & 0      \\
          &       & 0.9525  & 0.9528 \\
          &       & 1/4     & 1/4    \\
\hline \hline
\end{tabular}
\label{tab_Rh2S3_struc}
\end{center}
\end{table}

Rh$_3$S$_4$ has a complex monoclinic unit cell with 42 atoms, of which four Rh
and four S are inequivalent.
This structure is shown in Fig.~\ref{fig_cells}(b) and its geometrical 
properties are listed in Table \ref{tab_Rh3S4_struc} together with 
experimental values.
Two of the Rh atoms are still six-coordinated with S atoms forming a 
distorted octahedron.
However, the other two inequivalent Rh atoms form Rh$_6$ clusters surrounded
by S atoms.
In these clusters, the Rh-Rh bond lenghts are similar to the ones in pure
Rh.
Again, our DFT results show good agreement for lattice parameters
(with errors of 1.9\%, 2.1\%, and 1.8\%, for $a$, $b$, and $c$) and 
reproduce very well all the features of the atomic distribution in the
material.

\begin{table}
\caption{Lattice parameters, Wickoff atomic positions, and nearest-neighbor
         distances for Rh$_3$S$_4$ (space group 12, $C2/m$).
         The experimental data is from Ref.~\protect\onlinecite{Beck2000}.}
\begin{center}
\begin{tabular}{llcc}
\hline \hline
  & &  Experiment & Theory (this work) \\
\hline
    & $a$      & 10.29 \AA & 10.491 \AA \\
    & $b$      & 10.67 \AA & 10.899 \AA \\
    & $c$      & 6.212 \AA &  6.326 \AA \\
    & $\beta$  & 107.7$^{\circ}$ & 107.9$^{\circ}$ \\
\hline
    Rh(1) & $8j$  & 0.36278 & 0.3616 \\
          &       & 0.14491 & 0.1464 \\
          &       & 0.4501  & 0.4499 \\
    Rh(2) & $4i$  & 0.3509  & 0.3495 \\
          &       & 0       & 0      \\
          &       & 0.0546  & 0.0550 \\
    Rh(3) & $2c$  & 0       & 0      \\
          &       & 0       & 0      \\
          &       & 1/2     & 1/2    \\
    Rh(4) & $4g$  & 0       & 0      \\
          &       & 0.1596  & 0.1611 \\
          &       & 0       & 0      \\
    S(1)  & $4i$  & 0.4165  & 0.4157 \\
          &       & 0       & 0      \\
          &       & 0.7291  & 0.7256 \\
    S(2)  & $8j$  & 0.1262  & 0.1260 \\
          &       & 0.1582  & 0.1586 \\
          &       & 0.3854  & 0.3853 \\
    S(3)  & $4i$  & 0.1182  & 0.1171 \\
          &       & 0       & 0      \\
          &       & 0.8883  & 0.8883 \\
    S(4)  & $8j$  & 0.3493  & 0.3501 \\
          &       & 0.2111  & 0.2089 \\
          &       & 0.0990  & 0.1004 \\
\hline \hline
\end{tabular}
\label{tab_Rh3S4_struc}
\end{center}
\end{table}

As mentioned in the beginning of this subsection, the crystallographic 
analysis was unable to conclude whether Rh$_{17}$S$_{15}$ belongs to group
{\em Pm}\={3}{\em m} (\#221, $O_h$), {\em P}\=43{\em m} (\#215, $T_d$),
or {\em P}432 (\#207, $O$).\cite{Geller1962}
We have performed first-principles calculations starting from each of these
64-atom unit cell structures, including also atoms randomly displaced
from them.
In all cases the final relaxed state corresponds to the cubic 
{\em Pm}\={3}{\em m} case, so we suggest that this is indeed the ground
state of Rh$_{17}S_{15}$.
The structure and crystallographic data obtained are contained in
Fig.~\ref{fig_cells}(c) and Table~\ref{tab_Rh17S15_struc}.
The four inequivalent Rh atoms have different enviroments; 
the central one at $(1/2, 1/2, 1/2)$ is surrounded by six S atoms forming a
perfect octahedron, while
the other three Rh atoms have four S neighbours each.
The Rh-Rh bonds get stronger than in the previous two materials, being around 
3.5\% shorter than in the case of pure Rh.
Our DFT calculations reproduce all the experimental structural details, showing
a lattice parameter and bond lengths around 1\% larger.

\begin{table}
\caption{Lattice parameters, Wickoff atomic positions, and nearest-neighbor
         distances for Rh$_{17}$S$_{15}$ (space group 221, 
         {\em Pm}\={3}{\em m}).
         The experimental data is from Ref.~\protect\onlinecite{Geller1962}.}
\begin{center}
\begin{tabular}{llcc}
\hline \hline
  & & Experiment & Theory (this work) \\
\hline
  & $a$      &  9.911 \AA  &  10.025 \AA \\
\hline
    Rh(1) & $24m$ & 0.3564 & 0.3574 \\
          &       & 0.3564 & 0.3574 \\
          &       & 0.1435 & 0.1426 \\
    Rh(2) & $6e$  & 0      & 0      \\
          &       & 0      & 0      \\
          &       & 0.2338 & 0.2373 \\
    Rh(3) & $3d$  & 0      & 0      \\
          &       & 0      & 0      \\
          &       & 1/2    & 1/2    \\
    Rh(4) & $1b$  & 1/2    & 1/2    \\
          &       & 1/2    & 1/2    \\
          &       & 1/2    & 1/2    \\
    S(1)  & $12j$ & 0.1696 & 0.1696 \\
          &       & 0.1696 & 0.1696 \\
          &       & 1/2    & 1/2    \\
    S(2)  & $12i$ & 0.2310 & 0.2293 \\
          &       & 0.2310 & 0.2293 \\
          &       & 0      & 0      \\
    S(3)  & $6f$  & 1/2    & 1/2    \\
          &       & 1/2    & 1/2    \\
          &       & 0.2643 & 0.2630 \\
\hline \hline
\end{tabular}
\label{tab_Rh17S15_struc}
\end{center}
\end{table}

\subsubsection{Electronic properties}

\begin{figure}
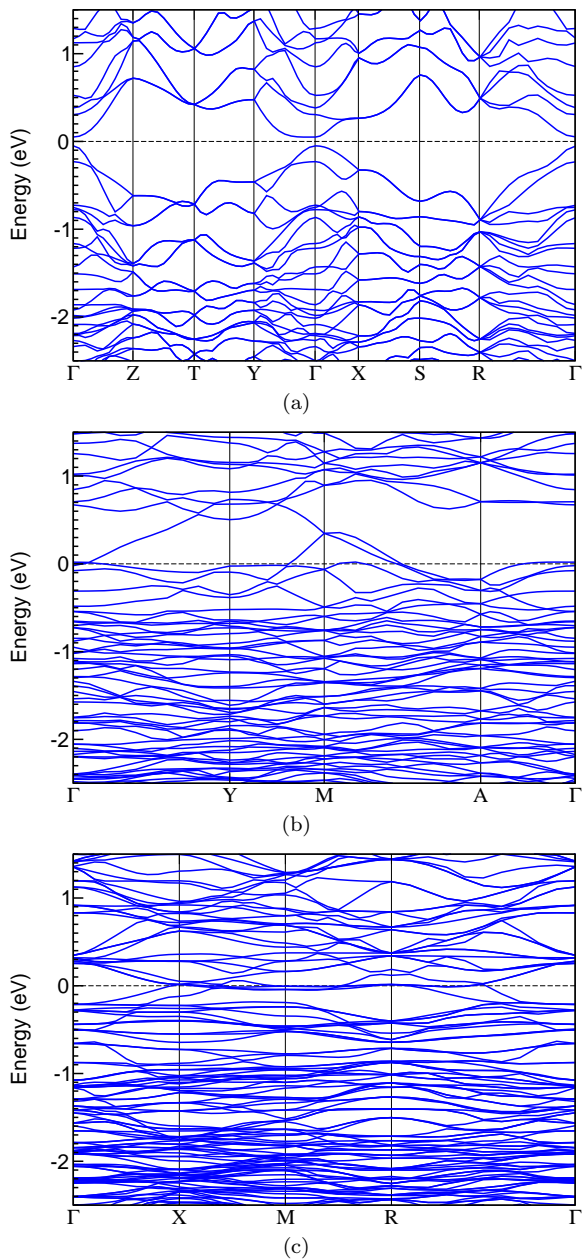

\subfigure[]{\includegraphics[width=3.0in]{bs_Rh2S3.eps}}
\subfigure[]{\includegraphics[width=3.0in]{bs_Rh3S4.eps}}
\subfigure[]{\includegraphics[width=3.0in]{bs_Rh17S15.eps}}
\caption{
Band structure for (a) Rh$_2$S$_3$, (b) Rh$_3$S$_4$,
and (c) Rh$_{17}$S$_{15}$. The Fermi level is at 0 eV.}
\label{fig_bandstructure}
\end{figure}

\begin{figure}
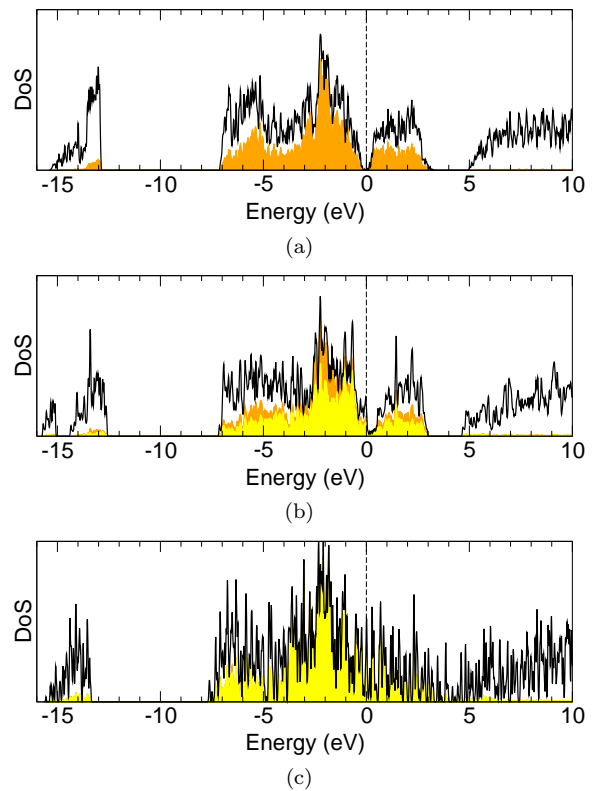

\subfigure[]{\includegraphics[width=3.0in]{dos_Rh2S3.eps}}
\subfigure[]{\includegraphics[width=3.0in]{dos_Rh3S4.eps}}
\subfigure[]{\includegraphics[width=3.0in]{dos_Rh17S15.eps}}
\caption{
Density of states for (a) Rh$_2$S$_3$, (b) Rh$_3$S$_4$,
and (c) Rh$_{17}$S$_{15}$. Black lines represent the total density of states,
and orange (yellow) shadow areas represent the density of states projected on 
the $d$ orbitals of the Rh atoms (the Rh atoms bonded to other Rh atoms).
The Fermi level is at 0 eV.}
\label{fig_dos}
\end{figure}

The band structures and densities of states computed using first-principles
calculations can be seen in Figs.~\ref{fig_bandstructure} and \ref{fig_dos}.
As the Rh content increases, the materials go from narrow-gap insulator
(Rh$_2$S$_3$) to metals (Rh$_3$S$_4$ and Rh$_{17}$S$_{15}$).

The band gap we obtained for Rh$_2$S$_3$ is 0.10 eV, in excellent agreement 
with the results of Raybaud and coworkers\cite{Raybaud1997b}, who obtained a
gap of 0.09 eV. Our density of states is also very similar to theirs; the 
$4d$ electrons of Rh are split in the $t_{2g}$ and $e_g$ manifolds by the
octahedral field, with the Fermi level laying in between.

The higher concentration of Rh in Rh$_3$S$_4$ pushes some of the Rh atoms to 
form bonds of similar length to the ones in the pure metal.
This causes the gap to close, being populated mainly by the $d$ states of 
the Rh clusters.
Our results show that the electronic analysis made by Beck and 
Hilbert \cite{Beck2000} using extended H\"uckel calculations is
quantitatively correct. 

In the case of Rh$_{17}$S$_{15}$, only one of the four inequivalent atoms
is surrounded by S atoms forming a (regular) octahedron.
There are strong bonds between the rest of Rh atoms, and the band
structure and density of states for this material shows its clear metallic
character.

\subsubsection{Formation enthalpy}

We have computed the cohesive energy of the three Rh-S materials by comparing
their energies in the ground state with the energy of the isolated atoms of
Rh and S.
Our results are: 4.73 eV/atom (Rh$_2$S$_3$), 4.83 eV/atom (Rh$_3$S$_4$), and
5.05 eV/atom (Rh$_{17}$S$_{15}$). 
The first of these is in good agreement with the experimental (4.61 eV/atom)
and theoretical (4.84 eV/atom) values reported in 
Ref.~\onlinecite{Raybaud1997}.

\begin{figure}
\includegraphics[width=2.5in]{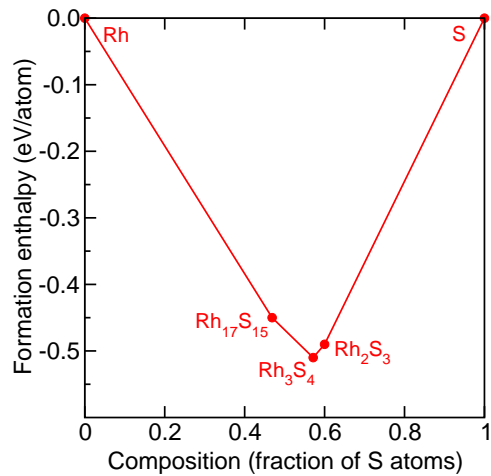}
\caption{Formation enthalpy for Rh-S materials.}
\label{fig_enthalpy}
\end{figure}

A useful reference for alloys is their enthalpy of formation, computed as the
difference between their cohesive energy and the (weighted) one of the pure 
elements that form them.
Figure \ref{fig_enthalpy} contains the formation enthalpy for each of the
rhodium sulfides, using $\alpha$-S and FCC Rh as the end components.
It shows a convex hull, as expected since the three materials are stable.
In fact, as mentioned in the introduction, the Rh-S electrocatalyst
commercially available contains a mixture of the three.\cite{Ziegelbauer2008} 

\section{Summary}

In this article we present a first-principles study of the three rhodium
sulfides that are positively known to exist. 
We have obtained the relaxed structures for these materials of interest for
catalysis.
For Rh$_2$S$_3$ and Rh$_3$S$_4$ the crystallographic structures we find are
very similar to the ones found experimentally, while for Rh$_{17}$S$_{15}$ we
identify one of the three configurations proposed experimentally as the atoms
adopt.
We have computed the band-structure and density of states for the three 
compounds, something that had only been done before for Rh$_2$S$_3$ using
first-principles.
We present too an analysis of the energetics of these chalcogenides, showing
the convex hull graph for these alloys.
As part of this study, we also performed what we believe is the first full
optimization of the 128-atom unit cell of $\alpha$-S using DFT, showing how
the PBE approximation gets
correctly the S-S bonds in the compound, but that it is necessary to include
a further correction to take into accound the van der Waals interactions 
between $S_8$ rings in the structure.


\acknowledgements
We thank J.\ M.\ Ziegelbauer and S.\ Mukerjee for valiable discussions, 
E.\ Canadell for a critical reading of the manuscript, 
and N.\ Singh-Miller for computational assistance.
We have used the {\sc XCrySDen}\cite{xcrysden} and {\sc Jmol}\cite{jmol}
graphics programs,
and the Bilbao Crystallographic Server.\cite{bilbao} 
We acknowledge support from the MURI Grant DAAD 19-03-1-0169.
O.\ D.\ acknowledges support from the {\em Ram\'on y Cajal} program of the
Spanish Government.



\end{document}